\documentclass[11pt,twoside]{article}
\usepackage{asp2004}
\usepackage{psfig}
\begin{document}
\title{Stellar Collisions and Black Hole Formation in Dense Star
 Clusters}

\author{Stephen L.\ W.\ McMillan}
\affil{Department of Physics, Drexel University, Philadelphia, PA
  19104, USA}
\author{Simon F. Portegies Zwart}
\affil{Astronomical Institute `Anton Pannekoek' and Institute for
  Computer Science, University of Amsterdam, Kruislaan 403, The
  Netherlands
}

\begin{abstract}
Close encounters and physical collisions between stars in young dense
clusters may lead to the formation of very massive stars and black
holes via runaway merging.  We examine critically some details of this
process, using N-body simulations and simple analytical estimates to
place limits on the cluster parameters for which it expected to occur.
For small clusters, the mass of the runaway is effectively limited by
the total number of high-mass stars in the system.  For sufficiently
dense larger clusters, the runaway mass is determined by the fraction
of stars that can mass segregate to the cluster core while still on
the main sequence.  The result is in the range commonly cited for
intermediate-mass black holes, such as that recently reported in the
Galactic center.
\end{abstract}

\section{Introduction}

The past decade has seen the discovery of a large number of massive
young star clusters throughout the local universe.  These systems are
large and young enough that they contain statistically significant
numbers of massive stars, affording us key insights into their initial
mass function and structural properties.  Of greatest interest to
dynamicists are those systems in which stellar dynamical time scales
are short enough that the cluster can undergo significant structural
change during the lifetimes of the most massive stars.  In such
clusters, dynamical evolution opens up entirely new avenues for
stellar and binary evolution, allowing the formation of stellar
species completely inaccessible by standard stellar and binary
evolutionary pathways.

From this perspective, the clusters listed by Portegies Zwart and
McMillan (these proceedings) represent an ideal combination of
properties, having ages of less than a few million years and
relaxation times of less than a few tens of millions of years.  In
these clusters, dynamical evolution, traditionally regarded as a
``slow'' process, actually occurs much more rapidly than the stellar
evolution of even the most massive stars.  Thus the dynamics controls
the early phases of these stars' lives.  We focus here on perhaps the
most dramatic environmental modification of standard stellar
evolution---repeated physical collisions between stars.  The scenario
described below is fast becoming the ``standard'' paradigm by which
runaway stellar mergers might occur, perhaps forming intermediate-mass
black holes (IMBHs) in sufficiently dense systems.

The possibility of multiple mergers of massive stars in dense stellar
systems was first demonstrated in N-body simulations of R\,136 by
Portegies Zwart et al.~(1999).  Subsequently Ebisuzaki et al.~(2001)
suggested that the ultraluminous X-ray source M82-X1 might be the
result of such a process; this possibility was explored in detail by
Portegies Zwart et al.~(2004).  Portegies Zwart \& McMillan (2002) and
McMillan \& Portegies Zwart (2003) have explored the possibility of
IMBH formation in young clusters in the Galactic Center.  Recently,
G\"urkan et al.~(2004) have reaffirmed the basic process using
Monte-Carlo simulations, and have carried out systematic studies of
runaway mergers in galactic nuclei.

In this paper we consider how a young star cluster might come to be in
such a high-density state, and look critically at the key physical
processes needed for a runaway merger to occur.  We then turn briefly
to the possibility of an IMBH in the center of the Milky Way Galaxy
(Maillard et al.~2004), showing how theory and recent observations may
provide a consistent picture of the Galactic center.


\section{Stellar Collisions}

Consider a massive object of mass $M=mM_\odot$ and radius $R=rR_\odot$
moving through a field of background stars of total mass density
$\rho=10^6\,\rho_6\,M_\odot/{\rm pc}^3$ and velocity dispersion
$v=10\,v_{10}\,{\rm km/s}$.  If $M$ and $R$ are large compared to the
masses and radii of other stars, and all velocities are small enough
that gravitational focusing dominates the total cross section, the
object's collision cross section is $\sigma \approx 2\pi G M R / v^2$,
nearly independent of the properties of the other stars.  The rate of
increase of the object's mass due to collisions then is
\begin{equation}
    \frac{dM}{dt} ~\approx~ \rho\sigma v
		  ~\approx~ 2\pi G M R \rho / v
		  ~=~ 6\times10^{-11}\, m \, r \,
    			\rho_6 \, v_{10}^{-1}~ M_\odot/{\rm yr}\,.
\end{equation}
If the object initially has $m=100\,m_{100}$ and we adopt a simple
mass--radius relation $r=3\,m^{1/2}$, then for the object to reach
$m\gg10^3$ in 3 Myr (to form an IMBH within the lifetime of a massive
star), the local density must satisfy
\begin{equation}
 	\rho_6 \ \ga\  350 ~ m_{100}^{-1/2}\ v_{10}
		 ~=~ \rho_{crit},\ \  {\rm say}\,.
\end{equation}
Such a density is much higher than the mean density of any known star
cluster, young or old.  For comparison, the average density of the
Arches cluster is $\rho_6\sim0.6$, that of a fairly compact globular
cluster is $\rho_6\sim0.01$, while even the most concentrated globular
cluster cores have $\rho_6\la 1-10$.

Mergers would be enhanced if the cluster were born very centrally
concentrated, as suggested by Portegies Zwart et al.~(2004) and
Merritt et al.~(2004).  As a simple limiting model of such a cluster,
consider the nearly isothermal system of total mass $M_c$ and
half-mass radius $r_h$, described by the density profile
\begin{eqnarray}
	\rho(r) &=& \frac{M_c}{8\pi r_h r^2}\,,\\
	M(r)    &=& {\textstyle\frac12}M_c\left(\frac{r}{r_h}\right)\,,
\end{eqnarray}
for $0 \le r \le 2r_h$.  Densities exceeding $\rho_{crit}$ are found
for $r<r_{crit}$, where
\begin{equation}
	r_{crit} ~=~ \sqrt{\frac{M_c}{8\pi r_h\rho_{crit}}}
		 ~=~ 2.3\times10^{-3} \  v_{10}^{1/2} m_{100}^{1/4}\ \ 
						 {\rm pc}\,,
\end{equation}
where the cluster velocity dispersion is $v = \sqrt{GM_c/2r_h}$.
However, the total mass contained within this radius is just
\begin{equation}
M_{crit} \approx 50\ v_{10}^{5/2} \ m_{100}^{1/4}\ M_\odot\,.
\end{equation}
We conclude that, for reasonable cluster parameters, there is too
little initial mass in the high-density region to accomplish the task
of forming a $\sim10^3\,M_\odot$ object in the time available.


\section{Cluster Dynamics}

Thus collisions in a static cluster core cannot lead to the formation
of an ultramassive object.  However, cluster dynamical evolution can
result in conditions much more favorable for a runaway merger to
occur.  The evolution of a cluster is governed by its half-mass
relaxation time, the time scale on which two-body encounters transport
energy around the system:
\begin{equation}
	t_{rh}\ \approx\ \frac{0.14\, M_c^{1/2} r_h^{3/2}}
				{G^{1/2} \langle m\rangle \ln\Lambda}
	      \ \approx\ 0.5\  v_{10}^3 \,/\, \bar{\rho}_6 \  {\rm Myr}
\label{trelax}
\end{equation}
(Heggie \& Hut 2003).  Here, $N$ is the number of stars in the system,
$\langle m\rangle = M_c/N$ is the mean stellar mass, taken here to be
$0.5M_\odot$, $\bar{\rho} = 3M_c/8\pi r_h^3 =
10^6\,\bar{\rho}_6\,M_\odot\,{\rm pc}^{-3}$ is the mean cluster
density, and $\ln\Lambda\sim \ln(0.1 N)\sim10$.  For an equal-mass
system, the time scale for dynamical evolution---the core collapse
time---is about $15 t_{rh}$, too long to cause significant structural
change within a few million years.  However, the presence of even a modest
range in masses greatly accelerates the process of core collapse
(Spitzer 1987).  The time scale for a star of mass $m$ to sink to the
cluster center as equipartition reduces its velocity is
\begin{equation}
      t_s(M) \ \sim\  \frac{\langle m\rangle}{m}\, t_r\,,\label{tseg}
\end{equation}
where $t_r\sim t_{rh}\,\bar{\rho}/\rho$ is the local relaxation time.

Portegies Zwart \& McMillan (2002) find that the most massive ($m \ga
20 M_\odot$) stars segregate rapidly to the cluster center, forming a
dense stellar subcore on a time scale $t_{cc}\sim0.2 t_{rh}$.  A
central density increase of 2--3 orders of magnitude is typical,
boosting even a relatively low-density core into the range where
collisions become common, and greatly increasing the supply of raw
material to form a collision runaway.  In systems with $t_{cc} \la 5$
Myr ($t_{rh}\la 25$ Myr), essentially all the massive stars in the
cluster reach the center before exploding as supernovae, and hence can
participate in the runaway process.  The Arches and Westerlund I fall
into this category; the Quintuplet, NGC 3603, and R\,136 all come
close.  In these cases, the maximum mass of the runaway is limited
primarily by the total number of massive stars in the
system---anywhere from a few percent to several tens of percent of the
total, depending on the cluster mass function.

In less dense or more massive clusters, the longer half-mass
relaxation time means that only a fraction of the massive stars
initially present in the system can reach the center in the time
available, but the total supply of mass may still ensure that a
runaway can occur.  We can estimate the amount of mass made available
by mass segregation as follows.  Again adopting an isothermal model to
simplify the calculation, and taking the segregation time scale from
Eq.~\ref{tseg}, it is easily shown that a star of mass $m$ can sink to
the center within time $T$ if its initial distance $r$ from the
cluster center satisfies
\begin{equation}
	r \la r_s(m) = r_h
		       \left(\frac{m}{\langle  m\rangle}\right)^{1/2}\,
		       \left(\frac{T}{t_{rh}}\right)^{1/2}\,.
\end{equation}
The fraction of stars of mass $m$ satisfying this relation is
\begin{equation}
	f(m) =	{\textstyle\frac12}
		\left(\frac{m}{\langle  m\rangle}\right)^{1/2}\,
		\left(\frac{T}{t_{rh}}\right)^{1/2}\,. 
\end{equation}
Choosing (again for simplicity) a mass function $dn/dm \sim m^{-2}$
for $0.1 M_\odot < m < 100 M_\odot$, we determine the total stellar
mass potentially available for mergers as
\begin{equation}
	M_s = \int_{m_{min}}^{100M_\odot}\,\frac{dn}{dm}\,m\,f(m)\ dm\,,
\end{equation}
where the lower mass limit $m_{min}$ is somewhat greater than the mean
stellar mass ($0.7\,M_\odot$ here), but is otherwise unimportant so
long as it is much less than the upper limit of $100\,M_\odot$.  The
result for the chosen mass function is
\begin{eqnarray}
 	M_s &=& 1.7 \left(\frac{T}{t_{rh}}\right)^{1/2} M_c\nonumber\\
	    &=& 1.5 \times 10^4 M_\odot\ v_{10}^{3/2}\ 
			\left(\frac{T}{3 {\rm\, Myr}}\right)^{1/2}\,,
				\nonumber
\end{eqnarray}
comfortably above the value needed for IMBH formation.

We have assumed here that the cluster relaxation time is long enough
that $f(m) \le 1$ for all $m$ of interest.  This is the case for
$t_{rh} \ga 50$ Myr.  An appropriately modified version of this
analysis for $t_{rh}<50$ Myr shows the available mass leveling off at
a fixed fraction of the cluster mass as $t_{rh}\rightarrow0$, as
expected.

\section{Mergers and Stellar Evolution}

Given that dynamical evolution can concentrate enough mass in a
cluster core for collisions to occur at a significant rate, we can
then ask (i) if the collisions actually lead to mergers, and (ii)
under what circumstances a runaway merger can occur.  The answer to
the first question is provided by the SPH simulations of Freitag \&
Benz (2001; see also Lombardi et al.~2003), who find that the low
relative velocities typical of these systems ensure that the colliding
stars usually merge with minimal mass loss.  In small systems
(containing less than a few tens of thousands of solar masses),
collision rates are significantly enhanced by the fact that the
massive object tends to form binaries, which are then perturbed into
eccentric orbits by encounters with other stars (Portegies Zwart \&
McMillan 2002).  Binary-induced mergers increase the collision cross
section, but they still require high central densities before the
(three-body) binary formation rates become significant.  In larger
systems, unbound collisions appear to be the norm.

Thus collisions naturally involve the most massive stars in the
cluster, and lead to the production of even more massive objects.  The
merger products are generally out of thermal equilibrium and often
rapidly rotating, with the result that their subsequent stellar
evolution is currently poorly understood.  In our simulations we
generally assume that the merged object evolves along a suitably
rejuvenated (non-rotating) track appropriate to its mass (Portegies
Zwart et al.~1997).  This is at best a crude approximation, but we
note that it probably underestimates the radius of the merger product
during the out-of-equilibrium phase and hence the likelihood of a
runaway.  Acting in the opposite sense is the fact that, while stellar
mass loss rates are very uncertain, ultramassive stars probably have
very strong winds.  Van Beveren (these proceedings) points out that if
the wind mass loss rate exceeds the accretion rate due to mergers,
then the entire runaway process may fail.  Our simulations generally
yield net merger accretion rates of $\sim10^{-3} M_\odot\,{\rm
yr}^{-1}$, suggesting that a high (but perhaps not impossibly so)
mass-loss rate is needed to shut the process down.

Of course, it must be conceded that next to nothing is known about the
detailed evolution and ultimate fate of stars hundreds or thousands of
times more massive than the Sun.  Nevertheless, the estimates
presented here make it clear that dynamical evolution in dense stellar
systems can easily produce conditions suitable for repeated stellar
collisions.  The collision runaway at the center of such a system
should be extremely luminous and eminently observable during its short
lifetime.  Observations of the cores of dense young star clusters in
our Galaxy and beyond may thus shed light on the structure and
lifetimes of such ultramassive stellar objects.


\section{An IMBH in the Galactic Center?}

\begin{figure}[htbp!]
~~~\psfig{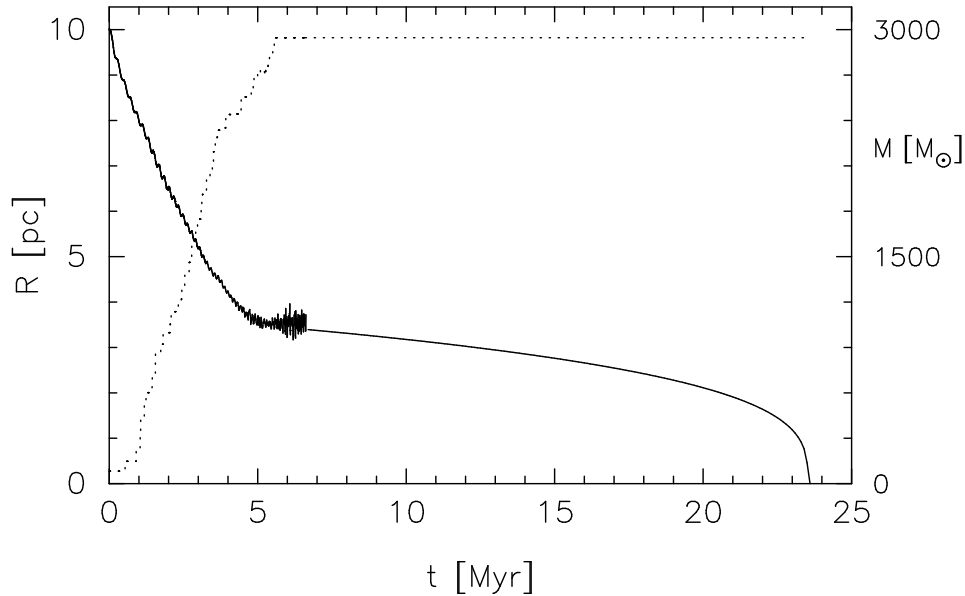}
\caption[]{Evolution of the Galactocentric radius $R$ (dotted line,
  left axis) and mass $M$ of the runaway merger product/IMBH (solid
  line, right axis) in a cluster of initial mass $4\times10^5 M_\odot$
  moving in the Galactic field.}\label{Fig1}
\end{figure}

If our evolving cluster happens to reside close to the Galactic
center, then dynamical friction will tend to drive it inward, raising
the possibility that mass segregation can lead to a collision runaway
en route, and that the resulting IMBH can be transported rapidly
toward the center by the much more massive cluster (McMillan \&
Portegies Zwart 2003).  Figure \ref{Fig1} illustrates this
possibility.  It shows the mass $M$ and Galactocentric distance $R$ of
the runaway formed in a cluster of initial mass $4\times10^5 M_\odot$,
placed in a circular orbit of radius 10 pc, as it spirals inward and
is disrupted by the Galactic field.  We clearly see the inward
transport of the growing merger product, terminating with the
dissolution of the cluster at $t\sim7$ Myr.  Subsequently, the black
hole sinks more slowly to the center, eventually reaching $R=0$ at
$\sim24$ Myr.  The early ($t\la7$ Myr) portion of the figure is from
an N-body simulation; the remainder is based on a simple analytical
model of the Galactic potential (Sanders \& Lowinger 1972).

The recent report by Maillard et al.~(2004) of a possible IMBH near
the Galactic center, at the heart of the swarm of stars known as
IRS\,13, provides the exciting prospect of confronting our models
directly with reality.  Portegies Zwart et al.~(2005, in preparation)
find that the observed properties of IRS\,13 are completely consistent
with theoretical expectations for a dense cluster remnant.  They
further estimate that the total IMBH mass resulting from runaway
collisions in the cluster population which produced the inner bulge is
consistent with the $\sim3\times10^6\,M_\odot$ supermassive black hole
at the Galactic center.


\section*{Acknowledgments}
This work was supported by NASA ATP grant NAG5-10775, the Royal
Netherlands Academy of Sciences (KNAW), the Dutch organization of
Science (NWO), and by the Netherlands Research School for Astronomy
(NOVA).


\end{document}